\documentstyle[12pt]{article}
 \textwidth
 16.5cm
 \oddsidemargin
 2.5cm
 \advance\oddsidemargin
 by
 -1in
 \evensidemargin
 0.0cm
 \advance\evensidemargin
 by
 -1in
 \marginparwidth
 1.9cm
 \marginparsep
 0.4cm
 \marginparpush
 0.4cm
 \topmargin
 -1.5cm
 \advance\topmargin
 by
 -0.0in
 \textheight
 22.0cm
 \makeindex
 \newcommand\noi{\noindent}
 \newcommand\beq{\begin{equation}}
 \newcommand\eeq{\end{equation}}
\newcommand{\doublespace}
 {
 \renewcommand{\baselinestretch}
 {1.6}
 \large\normalsize}

 \def\beqn{\begin{eqnarray}}
 \def\eeqn{\end{eqnarray}}
\begin{document}
\vspace*{.5cm}
\hspace*{9cm}{\Large MPI H-V3-1998}
%\\
%\hspace*{9.5cm}{\Large hep-ph/9607033}

\vspace*{2cm}
\begin{center}

{\Large{\bf
High-Energy
Polarimetry
at
RHIC}$\,$\footnote{Based on the talks presented by the author
at the Workshop on Polarimetry at RHIC,\\ RIKEN 
Research Center, BNL, July 20 - August 23, 1997}}

\vspace{1cm}

{\large
Boris~Z.~Kopeliovich}

\medskip

%{\sl
%Univesit\"at
%Heidelberg, Philosophenweg 19,
%69120 Heidelberg, Germany}
%\\
%
{\sl
Max-Planck-Institut
f\"ur
Kernphysik, Postfach 103980,
69029 Heidelberg, Germany}\\

{\sl
Joint Institute for Nuclear Research, Dubna,
141980 Moscow Region, Russia}

\end{center}

\vspace{1.5cm}
\begin{center}

{\large\sl This paper is dedicated to the memory of my teacher\\

\medskip

Professor Lev Lapidus}

\vspace{2.5cm}

{\large\bf
Abstract}\\ 
\end{center}

We compare a few types of high energy 
reactions which seem to be practical 
for polarimetry at RHIC. Coulomb-nuclear interference (CNI)
in $pp$ elastic scattering
leads to a nearly energy-independent left-right
asymmetry $A_N(t)$ at small $t$.
The systematical uncertainty of this method is evaluated 
to be $\sim 10\%$.

The CNI in proton-nucleus elastic scattering 
is predicted to result in 
larger values of $A_N(t)$ and occurs at larger momentum transfer
than in $pp$ elastic scattering. This energy independent 
asymmetry can be used for the
polarimetry.

As an absolute polarimeter one can 
use elastic $pp$ scattering
on a fixed target
at large $|t| \sim 1 - 1.5\ GeV^2$, where $A_N(t)$ is 
reasonably large and nearly energy
independent. Although it cannot be reliably calculated, one
can calibrate the polarimeter by measuring the polarisation of 
the recoil protons.

\doublespace

\newpage

\noi
{\large\bf
1. Introduction}\\

High-energy polarised proton
beams
are
under
construction at RHIC and a
wide
program
of
polarisation phenomena
study
is
planned. However, one faces
the
problem
of
measurement of the polarisation
of
the
beams.
A prospective polarimeter is supposed
to be able to
provide a fast (minutes/hours)
measurement
of
the
polarisation in a wide energy range
($25
-
250\
GeV$) with a statistical
error
within
$5\%$
and a systematical uncertainty of
the
same
order \cite{yousef}.

The
single
asymmetry
is believed to vanish
at
high
energies,
what is, however, not true
in
some
cases
which
can
be
used
for
high-energy
polarimetry.

The main problem is lack
of
reliable
and
accurate data on analysing power
$A_N$
of
hadronic
reactions at high energy. One
can
either
use
available data on analysing
power,
or
to
measure it in the same
experiment. In
some
cases
$A_N$ can be
reasonably
well
predicted
theoretically.

This
note is not a review of the present status of high
energy
polarimetry,
but includes only a few examples 
of polarimeters which seem to
be practical
for
RHIC.

\vspace{1cm}

\noi
{\large\bf
2. Pion
polarimeter}\\

Polarisation
in
inclusive
reactions
at
fixed
Feynman
$x_F$
is expected to be
nearly
energy
independent
(Feynman scaling) in
the
high-energy
limit.
In some
cases
it
reaches
a few tens percent at large
$p_T$
and
$x_F$.
As
an
example,
reaction of
inclusive
pion
production,
\beq
p + p \to
\pi^{\pm}
+
X
\label{1}
\eeq
\noi
exhibits
these
features
\cite{e704-pi}. Therefore,
this
reaction
is a very good
candidate
\cite{yousef}
for polarimetry at RHIC, provided
that
its
analysing power
is
known. Unfortunately,
the measurement of analysing
power
$A_N(p_T,x_F)$
of (\ref{1}) was performed at high energies
only once 
\cite{e704-pi}, at 
energy
$E=200\
GeV$ and with
a proton
target.
The polarimeter,
however,
is
supposed
to work in the full beam energy range
from
AGS
to the 
maximal energy of RHIC
$E_{max}\approx
250\
GeV$
in the lab frame. Therefore, 
energy
independence
of $A_N$
is
assumed in \cite{yousef}.
Moreover, only a carbon target is
feasible
to
be
used at RHIC.  Lacking
data
for
reaction
(\ref{1}) on a carbon one
is enforced either 
to assume no $A$-independence
of the
analysing
power \cite{yousef}, or to
measure it
with low energy beam of
known polarisation 
and assume that the analysing power does
not change
through the whole energy range
of RHIC.

We are going to look more
attentively
at
these
assumptions
and
their
justification.

\bigskip

\noi
{\bf
2.1 A-dependence}\\

$A$-dependence of
inclusive
particle
production
at large transverse momentum is known
to
exhibit
the
so
called Cronin effect \cite{cronin},
namely,
the
exponent
$\alpha$ describing
effectively
the
$A$-dependence
($A^{\alpha}$) exceeds one,
{\it
i.e.}
the
inclusive production rate on a
nucleus
is
more
than $A$ times larger than that on a
free
nucleon.
It looks like
the bound nucleons help each
other
producing
the
high-$p_T$
particle. Although
no
satisfactory numerical explanation
of
this
effect
is still known, the
source
the enhancement
is
well
understood on a qualitative
level. The
projectile
partons
experience
multiple interactions in the
nucleus,
and
the
higher the $p_T$ is, the larger
is
the
mean
number of rescatterings. This
is
because
the
large $p_T$ is distributed
over
many
interactions with smaller
momentum transfer. The multiple interactions
lead
to
a
steeper $A$-dependence,
since
each
interactions
adds a factor $\sim A^{1/3}$
due
to
integration
over the longitudinal position
of
the
interaction
point. Thus, the enhanced
$A$-dependence
is
a
clear signal
of
multiple
rescattering. A
model-independent relation
between
the
exponent
$\alpha$ and the mean number
of
rescatterings
is
derived
in
\cite{dijet}\footnote{Compared to \cite{dijet}
Eq.~(\ref{2})
includes
one additional interaction which triggers
the
process
(\ref{1}).},
\beq
\langle
n\rangle
=
3\alpha
-
1
+
\sigma_0\langle
T\rangle
\label{2}
\eeq
\noi
Here $\sigma_0 = (\int dk^2\
d\sigma/dk^2_T)$
is
the
total parton -
nucleon
cross
section.
$\langle T\rangle$ is
the
mean
nuclear
thickness,
\beq
\langle
T\rangle
=
\frac{1}{A}\
\int
d^2b\
T^2(b)\
,
\label{4}
\eeq
\noi
where
\beq
T(b)
=
\int\limits_{-\infty}^{\infty}dz\
\rho_A(b,z)
\label{5}
\eeq
\noi
is the nuclear thickness
function
at
impact
parameter $b$. The
nuclear
density
$\rho_A(b,z)$
depends on $b$ and
longitudinal
coordinate
$z$. 

Data \cite{antreasyan} on pion
inclusive
production
show
that at $p_T = 1\ GeV$
the
exponent
$\alpha
\approx
0.9$.
The mean nuclear thickness of
carbon
with
realistic
nuclear density \cite{ws} is
$\langle
T\rangle_C
=
0.33\ fm^{-2}$.
According to (\ref{2})
the
mean
number
of parton rescatterings
in
carbon
is
$\langle n\rangle_C
=
2$.

One comes to a
similar
result in a different way 
comparing the 
probabilities of single
and
double
rescatterings.
\beq
\frac{W_2(p_T)}{W_1(p_T)}
=
{1\over 2}\
\sigma_0\
\langle
T\rangle\
\exp(p_T^2B_{qN}/2)
\label{3a}
\eeq
\noi

For a rough estimate we can use
the
quark -
nucleon
interaction
cross
section
given by
the
constituent
quark
model $\sigma_0
\sim
10\
mb$.
We assume in (\ref{5})
a
Gaussian
$p_T$-dependence
of the quark-nucleon
inclusive
cross
section.
The slope parameter of
quark-quark
scattering
is
related to the mean radius
of
the
constituent
quark \cite{povh} $R_q^2 \approx
0.44\
fm^2$.
Correspondingly,
$B_{qN}\approx
R_q^2/3
\approx
3.5\
GeV^{-2}$.
Using (\ref{3a}) we can
estimate
the
ratio
of double to single
scattering
contributions
at
$(W_2/W_1)_C \approx 1$ at $p_T^2
=
1\
GeV^2$,
what agrees reasonably well with
the
previous
evaluation
(\ref{2}).

Neglecting the higher that
double
rescattering terms we can
estimate
the
analysing power of
reaction (\ref{1}) on a
nuclear target. Since
the
projectile quark interacts
incoherently
with
the
nucleus at large $p_T$,
one
can
write,
\beqn
A_N^{(A)}(p_T) &=&
\left[1
+
\frac{W_2(p_T)}
{W_1(p_T)}\right]^{-1}
\left\{A_N^{(N)}(p_T)
+
\frac{2}{\pi}
\sigma_{qN}
B_{qN}
\exp(B_{qN}p_T^2)\langle
T\rangle\
\right.\nonumber\\
&\times&\left.
\int
d^2k_T\
A_N^{(N)}(k_T)\exp(-B_{qN}k_T^2)
\exp\left[-B_{qN}(\vec p_T
-
\vec
k_T)^2\right]
\right\}
\label{6}
\eeqn
\noi
This formula can be used
to
predict
$A$-dependence
of polarisation effects in
inclusive
reactions
on moderately heavy nuclei and at
moderately high $p_T$ (otherwise
higher
order
multiple scattering terms are important).
This
is
not, however, our present objective.  We
need
just
a rough estimate to check whether
one can expect a weak or a strong
$A$-dependence.

The single asymmetry
$A_N^{(N)}(p_T)$
in
inclusive
pion production off a free proton
as
measured
in
the E704 experiment \cite{e704-pi} is
nearly
zero
at
$p_T < 0.5\ GeV$, linearly
grows
with
$p_T$
up to $p_T \approx 1\ GeV$,
and
is
approximately
constant at higher $p_T$. For
an
estimate
we
fix $A_N^{(N)}(k_T)$ in the
second
term
in
(\ref{6}) at the mean
value
of
momentum
transfer in each of the
double
scatterings
$k_T^2
= p_T^2/2$. Then the asymmetry
in
pion
production
off carbon at $p_T = 1\ GeV$ is
expected
to
be
$A_N^{(C)}(p_T=1\
GeV)
\approx
\frac{1}{2}A_N^{(N)}$.
Thus, we expect quite a strong $A$-dependence of the 
analysing power of reaction (\ref{1}).

\bigskip

\noi
{\bf
2.2
Energy
dependence}\\

At high
Feynman
$x_F$
the
cross section of an inclusive
reaction
can
be
described by a
triple-Regge
graph. An
example
for $pN \to \pi^+X$ is
shown
in
Fig.~1.

\begin{figure}[tbh]
\includegraphics{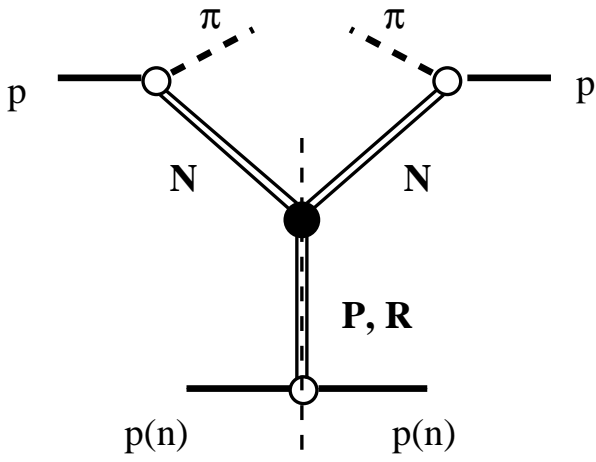}
\begin{center}
\vspace{6cm}
\parbox{13cm}
 {\caption[Delta]
 {\it The triple-Pomeron graph for the cross
 section of the reaction $pN \to
 \pi X$.  The thin dashed line shows
 that
 only the absorptive part of the
 amplitude is included. The upper legs are
the nucleon Reggeons, the bottom one is either
the Pomeron or the leading Reggeons, 
$\omega,\ f,\ \rho,\ a_2$.}
\label{fig1}}
\end{center}
\end{figure}

The bottom leg of this graph can
be
either
the
Pomeron, or a Reggeon ($f,\
\omega,\
\rho,\
a_2$).
In the latter case the
inclusive
cross
section at fixed $x_F$
decreases
$\propto
1/\sqrt{s}$.
However,
this
bottom part of the graph can be
treated
as
an
absorptive part of the
amplitude
of
elastic
scattering of the $N$-Reggeon
(upper
legs
in
Fig.~1) on the target. This
amplitude
is
subject
to exchange
degeneracy,
therefore,
the
Reggeons
should
cancel
each
other in the imaginary part.
Nevertheless, exchange degeneracy is
known
to
be
broken in $pp$ total cross
section. This
is
why
$\sigma^{pp}_{tot}(E)$ decreases
at
low
and
moderate energies.  To
evaluate
the
energy
dependence we can use the
the
data
on
$\sigma^{pp}_{tot}(E)$.
The
energy $E$ should be taken
at
smaller
value
than the energy $E_{pp}$
of
reaction
(\ref{1}):
\beq
E =
(1
-
x_F)E_{pp}
\label{7}
\eeq
For example, a variation of the
beam
energy
$E_{pp}$
in reaction (\ref{1}) from $25\ GeV$
to
$250\
GeV$ at $x_F
=
0.8$
corresponds to the
energy
$E$
variation
for $\sigma^{pp}_{tot}(E)$ from $5\
GeV$
to
$50\
GeV$.  The
cross
section
$\sigma^{pp}_{tot}(E)$
decreases in this interval by
nearly
$10\%$.
This
estimate shows that the energy
dependence
can
be
substantial.

Another source of energy dependence
is
a
strong
variation of the $A$-dependence
of
the
inclusive
cross section at high
$p_T$
with
energy
\cite{antreasyan}. It leads
to
variation
of the cross section on carbon by
$30\%$
in
the energy interval $200 - 400\ GeV$. We
would
expect even 
more substantial change in the RHIC
energy
range.
This means,
particularly,
that
even
if the $A$-dependence is known at
one
energy,
one
cannot assume it to be the same
in
the
whole
energy
range.

\bigskip

\noi
{\bf
2.3
Isospin
dependence}\\

Another source of nuclear
dependence
of
the
asymmetry is a possible
difference
between
proton
and neutron targets.  The
triple-Reggeon
graph
in
Fig.~1 is sensitive to the isospin
of
the
target
if the bottom leg of this graph
is
an
isospin
vector Reggeon ($\rho,\
a_2$). As
was
mentioned
above, the exchange degeneracy
leads
to
a
substantial cancellation between
$\rho$
and
$a_2$.
These Reggeons contribute to
the
absorptive
part
of the $N$-Reggeon - nucleon
amplitude
much
less
than the dominant $\omega$
and
$f$
Reggeons.
Therefore, we do not expect
a
strong
isospin
dependence of the asymmetry.
To
evaluate
the
effect we can use the
known
difference
between
$\sigma^{pp}_{tot}(E)$
and
$\sigma^{pn}_{tot}(E)$.
The energy $E$ should
be
taken
at a 
smaller value (\ref{7}).  
\vspace{1cm}

\noi
{\large\bf 3. Coulomb
-
nuclear
interference
(CNI)}\\

This method of polarimetry
has
been
under
discussion since 1974 when a
nearly
energy
independent asymmetry due
to
interference
between electromagnetic and hadronic
amplitude
was
claimed
in \cite{kl,bgl}.
Assuming the hadronic amplitude 
to be spin-independent at high energies one
can predict the single spin
asymmetry
$A_N(t)$
in elastic $pp$ scattering,
which behaves at small $t$ as \cite{kl}, 
\beq
A^{pp}_N(t) = A^{pp}_N(t_p)
\frac{4y^{3/2}}{3y^2 + 1}\ ,
\label{7a}
\eeq
\noi
where $y=|t|/t_p$ and $t_p = 
8\sqrt{3}\pi\alpha/\sigma^{pp}_{tot}$ is
the value of $|t|$ where the asymmetry has 
a maximum value,
\beq
A^{pp}_N(t_p) = 
\frac{\sqrt{3t_p}}{4m_p}
(\mu_p - 1)\ ,
\label{7b}
\eeq
\noi
with the proton magnetic moment $\mu_p$.
We neglect 
in (\ref{7b}) the real part of the hadronic amplitude 
and the Bethe phase. Both can be easily incorporated in
this formula \cite{bgl}.

The first measurements of $A^{pp}_N(t)$ 
by the E704 collaboration
\cite{e704} at $200\ GeV$ confirmed prediction \cite{kl}. 
The data are compared with
(\ref{7b}) in Fig.~2.

\begin{figure}[tbh]
\includegraphics{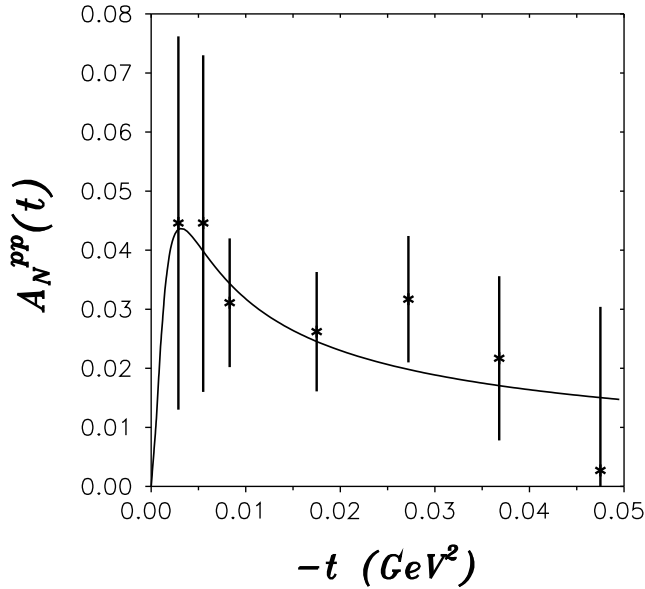}
\begin{center}
\vspace{7.5cm}
\parbox{13cm}
 {\caption[Delta]
 {\it Data \cite{e704,e704-c} for the asymmetry in
polarised elastic $pp$ scattering in the CNI region
at $200\ GeV$. The curve predicted in \cite{kl}
corresponds to (\ref{7a}).}
\label{fig1}}
\end{center}
\end{figure}

The CNI asymmetry is nearly
energy
independent
(it
depends only on $\sigma^{pp}_{tot}$
and
the
ratio
of real to imaginary parts of
the
forward
elastic
amplitude, which are pretty well
known
in
the
energy range under discussion) and
can
serve
for
polarimetry at high energies.

It
was
noticed,
however, in\cite{kz,larry} that
presence
of
a
spin-flip component of the
hadronic
amplitude
affects
the
value of CNI
asymmetry.
The relative deviation of 
$A_N$ from the nominal value (\ref{7b}) is
$-2{\rm IM}\,r_5/(\mu - 1)$ \cite{kl,bgl}, 
where $r_5
=(2m/\sqrt{-t})
\phi_5/{\rm Im}(\phi_1+\phi_3)$\footnote{According to our definition
$r_5$ is twice as small as the anomalous 
magnetic moment of the Pomeron $\mu_P$ introduced in \cite{kz}.
Our definition of $r_5$ is also different 
by $90^0$ phase from $\tau$ used
in \cite{larry}}.
This might be a 
substantial correction
to the predicted asymmetry (\ref{7b}).

Such a
sensitivity to the unknown
spin-flip
component
of
the Pomeron gives an unique
opportunity
to
measure
it \cite{kz} provided that
the
beam
polarisation
is known.  However, it brings
an
uncertainty
to
the
CNI
polarimetry
\cite{larry}.\\

\bigskip

\noi
{\bf 3.1 What do we know about the hadronic spin-flip
at high energy and small
$t$?}\\

Even in this circumstances the situation with
CNI
polarimetry is not hopeless.  
One can find solid
arguments, both experimental and
theoretical,
reducing the uncertainty down to a few
percent
\cite{workshop}.

\begin{itemize} \item The results of the E704
experiment \cite{e704} depicted in Fig.~2
show no deviation within the (quite large)
error bars from the prediction
\cite{kl,bgl}
based on a spinless hadronic amplitude.
Therefore
the data impose an upper limit on a
possible
hadronic spin-flip.  We cannot expect
a
substantial real part of this amplitude at
high
energy, otherwise it would interfere with
the
imaginary non-flip part resulting in a
large
polarisation in $pp$ elastic
scattering,
contradicting the data (see in
\cite{workshop}). If
spin-flip amplitude $\phi_5$ is imaginary,
the
analysis \cite{nigel,larry} of data \cite{e704} leads
to
a restriction ${\rm Im}\,r_5 < 0.15$. 

\item
The data on asymmetry in $\pi^+p$ and
$\pi^-p$
elastic scattering, if they are summed, may
have
contribution only from the
Reggeons with even signature,
{\sl i.e.} from the interference between the
Pomeron
and the $f$-Reggeon.
The upper limit on the Pomeron spin-flip
component corresponds to a pure non-flip
$f$-Reggeon
(provided that spin-flip to non-flip ratios
for
$f$ and $P$ have the same sign, what looks
very
natural, for instance in the model of
the
$f$-dominated Pomeron).  The analysis of
available
data in the energy range $6 - 14\ GeV$ performed
in
\cite{k80,workshop} leads (assuming Regge factorisation) 
to a restriction $r_5 <
0.1$, which is in agreement with the
above estimate. 

Although this analyses is performed at 
quite low energies
(available data at high energies are
not
sufficiently precise), we do not expect
a
substantial energy dependence of $r_5$.
According
to the recent results from HERA for the
proton
structure function $F_2(x,Q^2)$, the steepness
of
energy (or $1/x$) dependence of $F_2$
is
controlled by the mean size ($1/Q^2$) of
the
hadronic fluctuations of the virtual photon:
the
smaller the size is, the steeper is the growth
of $F_2$. It
is found in \cite{z,kz} (see below) that
the
spin-flip part of the amplitude originates from
a
smaller size configurations of the proton than
the
non-flip part. Therefore, one can expect a
rising
energy dependence of $r_5$.  However, even a
most
extreme evaluation of this effect,
assuming a scale of $10\ GeV^2$,
results in a growth of ${\rm Im}/,r_5$ from $E=10\ GeV$ to
$250\
GeV$ by only
$30\%$.

\item
A model independent (although with some approximations)
amplitude analyses \cite{ira} of pion-proton 
elastic and charge-exchange scattering data
shows that the energy-independent part of the iso-singlet
(in t-channel) spin-flip amplitude corresponds 
to ${\rm Im}\,r_5$, which does not exceed $15\%$. This analysis,
however, overlaps with the one we discussed previously, 
since it is essentially based on
the same data.

\item
A perturbative QCD evaluation of the
Pomeron
spin-flip amplitude was done in \cite{z,kz}.
It
is widely believed that the perturbative
Pomeron
contains no spin-flip component because
the
quark-gluon vertex conserves
helicity. However,
the proton helicity differs from the sum of
the
helicities of the quarks, because those
have
transverse motion, {\sl i.e.} their momenta
are
not parallel to the proton one.  This fact
can
lead to a nonzero proton spin-flip.
Calculations
performed in the nonrelativistic constituent
quark
model (in the Breit frame) \cite{kz}
show,
however, that these corrections cancel if
the
proton has a symmetric quark structure.  Only
in
the case when a component with a compact $qq$
pair
(diquark) is enhanced in the proton wave
function,
the Pomeron-proton vertex acquires a
nonzero
spin-flip part.  The smaller the mean
diquark
radius $r_D$ is, the larger is $r_5$.
For
reasonable values of $r_D > 0.2\ fm$ the spin
flip
fraction ${\rm Im}\,r_5$ ranges within $10\%$ in the
CNI
region of momentum
transfer.

Although oversimplified, these
calculations
demonstrate that within perturbative QCD there
is
no source of a large spin-flip component of
the
Pomeron.  However, the non-perturbative
effects might be a potential source 
of a spin-flip
amplitude.

\item
One can effectively include
the non-perturbative
effects switching to hadronic
representation,
which should be equivalent to the QCD
treatment due to quark-hadron duality.
 The Reggeon-proton vertex is
decomposed over hadronic states in
\cite{itep} (see also \cite{gol})
using a two-pion approximation for
the
$t$-channel exchange and a two-component
($N,\
\Delta$) intermediate state for the nucleon.
It
turns out that the resulting proton -
Reggeon
spin-flip vertex essentially correlates with
the
isospin in $t$-channel \cite{gol}.  Namely, the
contributions
of intermediate $N$ and $\Delta$ nearly
cancel
each other for the isosinglet Reggeons ($P,\
f,\
\omega$) resulting in ${\rm Im}\,
r_5 \approx 0.05$.  On the
other hand, $N$ and $\Delta$ add up in
the
isovector amplitude ($\rho,\ a_2$) leading to
an
order of magnitude larger value of
$r_5$.

This approach 
includes effectively the 
non-perturbative QCD effects and is based
on
completely different approximations than
the
perturbative calculations
\cite{z,kz}.
Nevertheless, it leads to a similar evaluation
for
the Pomeron $r_5$.  It makes these
theoretical
expectations quite
convincing.

\end{itemize}

Summarizing our experimental and
theoretical
knowledge of the hadronic spin-flip at high
energy
and small $t$ we conclude that the
systematical
uncertainty of CNI polarimetry does not
exceed
$10\%$.

Note that CNI method can be used both in
the collider
and fixed target experiments.  Provided that
a
polarised proton jet target will be available
one
can calibrate the CNI polarimeter, {\it
i.e.}
eliminate the uncertainty related to the
hadronic
spin-flip component.  This is another
potential
advantage of the CNI polarimeter compared to
one based on 
inclusive pion production. Indeed, it can be calibrated on
a
polarised target only if the pion is detected
in
 reversed kinematics, {\it i.e.} in
the
fragmentation region of the polarised
target, what is difficult to do.

\bigskip

{\bf 3.2 CNI on nuclear
targets}\\

The spin structure of the elastic proton-nucleus scattering
is simpler than that in $pp$. There are only two
spin amplitudes if the nucleus is spinless, otherwise
other amplitudes are suppressed by factor $1/A$. 
Besides, the main contribution to the $pp$ spin-flip amplitude
$\phi_5$, which comes mainly from the iso-vector Reggeons $\rho$
and $a_2$, are either forbidden or suppressed by $1/A$.
Thus, the uncertainty of CNI polarimetry may be 
substantially reduced.

The $t$ - dependence of single asymmetry
in
polarised elastic $p - A$ scattering in the
CNI
region is similar to that in $pp$ scattering.
The
position $t_p$ of the maximum of the asymmetry
and
the value of $A_N(t_p)$ 
are controlled by
the
magnitude of the $\sigma_{tot}^{pA}$ and
the electric 
charge of the
nucleus $Z$ (compare with (\ref{7a})-(\ref{7b})),
\beq
t^{pA}_p = \frac{Z\sigma_{tot}^{pp}}
{\sigma_{tot}^{pA}}t^{pp}_p\ .
\label{8}
\eeq
\noi
Correspondingly,
\beq
A^{pA}_N(t^{pA}_p) = 
\sqrt{\frac{Z\sigma_{tot}^{pp}}
{\sigma_{tot}^{pA}}}A^{pp}_N(t^{pp}_p)\ .
\label{9}
\eeq
It turns out that the position of the maximum of 
$A^{pA}_N$ and its value are not much different from 
that in $pp$ scattering.

It is proved in \cite{k-bnl} that if  
$r_5$ is imaginary, it is independent of $A$,
{\it i.e.} is the
same as the iso-singlet part or
$r_5$ in $pp$ elastic scattering.
Therefore,
the hadronic spin-flip brings the same
uncertainty
to CNI polarimetry on nuclear targets.

The differential cross section of proton -
nucleus
elastic scattering exhibits a
diffractive
structure, {\sl i.e.} series of maxima and
minima
\cite{schiz}. This is known to be a result
of
destructive interference between different
terms
in the multiple scattering
series. Imaginary
part of the elastic amplitude changes sign
at
positions of the minima. The first minimum
happens
at $|t| \sim 3/R_A^2$, which is in the
CNI
region and for heavy nuclei is quite close
to
$|t_p|$ given by
(\ref{9}).

As soon as the imaginary part of the
hadronic component of the non-flip elastic
amplitude changes sign, the CNI asymmetry does
the
same. We expect a nontrivial behaviour of $A_N(t)$
in the vicinity of the minimum of the differential
cross section, what resembles
the $pp$ elastic scattering at much larger $t$ 
(see the next section). Namely,
$A_N(t)$
reaches a sharp positive maximum, then
changes
sign and develops a sharp negative
minimum.

Indeed, the CNI asymmetry is given by expression,
\beq
A^{pA}_N(s,t)\left(\frac{d\sigma^{pA}_{el}}{dt}\right) = 
\frac{Z\alpha\sigma_{tot}^{pA}}
{2m_pq}\ F_A^C(q^2)F_A^H(q^2)
\left[\mu_p-1-2{\rm Im}\,r_5\right]\ ,
\label{10}
\eeq
\noi
where $t = - q^2$, and $\vec q$ is the 
transverse momentum transfer,
\beq
\sigma_{tot}^{pA} = 
2\int d^2b \left[
1 - e^{-{1\over 2}\sigma_{tot}^{pN}T(b)}\ .
\right]
\label{11}
\eeq
\noi
Here $T(b)$ 
is the nuclear thickness function defined in (\ref{5}).

The electromagnetic and hadronic nuclear formfactors in
(\ref{10}) read,
\beq
F_A^C(q^2) = \frac{1}{A}
\int d^2b \ e^{i\vec q\vec b} T(b)\ ,
\label{13}
\eeq
\beq
F_A^H(q^2) = \frac{1}{2\sigma_{tot}^{pA}}
\int d^2b \ e^{i\vec q\vec b} \left[
1 - e^{-{1\over 2}\sigma_{tot}^{pN}T(b)}
\right]\ .
\label{14}
\eeq
The elastic $pA$ differential cross section reads,
\beq
\frac{d\sigma^{pA}_{el}}{dt} = 
\frac{
\left[\sigma^{pA}_{tot}
F^H_A(t)\right]^2}{16\pi} + 
4\pi\left(\frac{Z\alpha F^C_A(t)}{t}\right)^2
\label{15}
\eeq
\noi
We neglect the ratio of real to imaginary parts of the 
$pA$ elastic amplitude, which is smaller than that 
in $pp$ scattering. The Bethe phase is neglected as well, 
although it might be a substantial correction for heavy nuclei 
\cite{nigel}. These corrections are to be done \cite{k-bnl} for 
a precise prediction,
but we can neglect them to demonstrate the magnitude of
the polarisation effects.

Our predictions for the CNI contribution to the 
single asymmetry for elastic scattering
of polarised protons on carbon and lead 
are shown in Fig.~3. As we expected, the dip structures
in the differential cross section reflect in a
nontrivial $t$-dependence of $A_N(t)$.

\begin{figure}[tbh]
\includegraphics{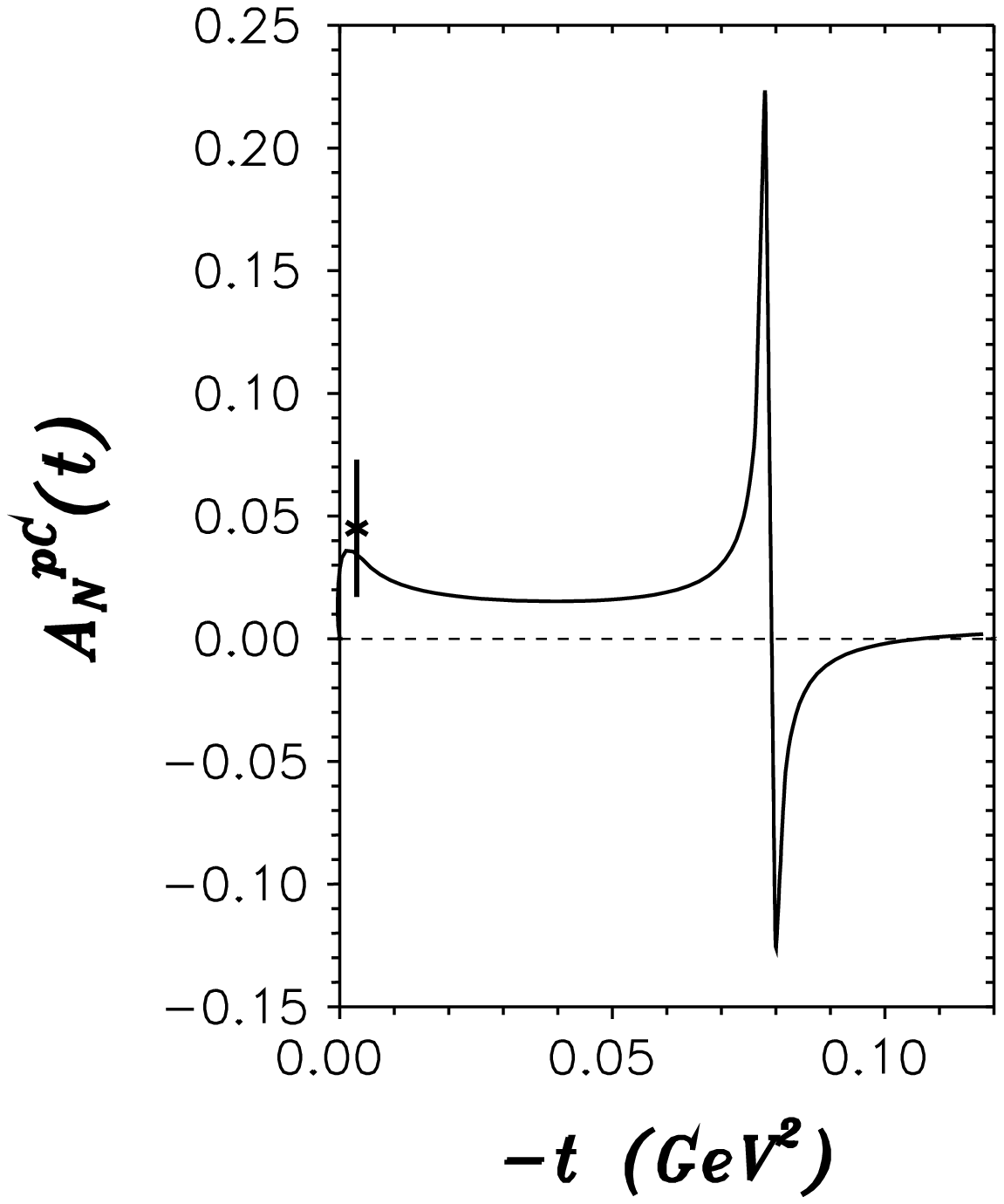}
\includegraphics{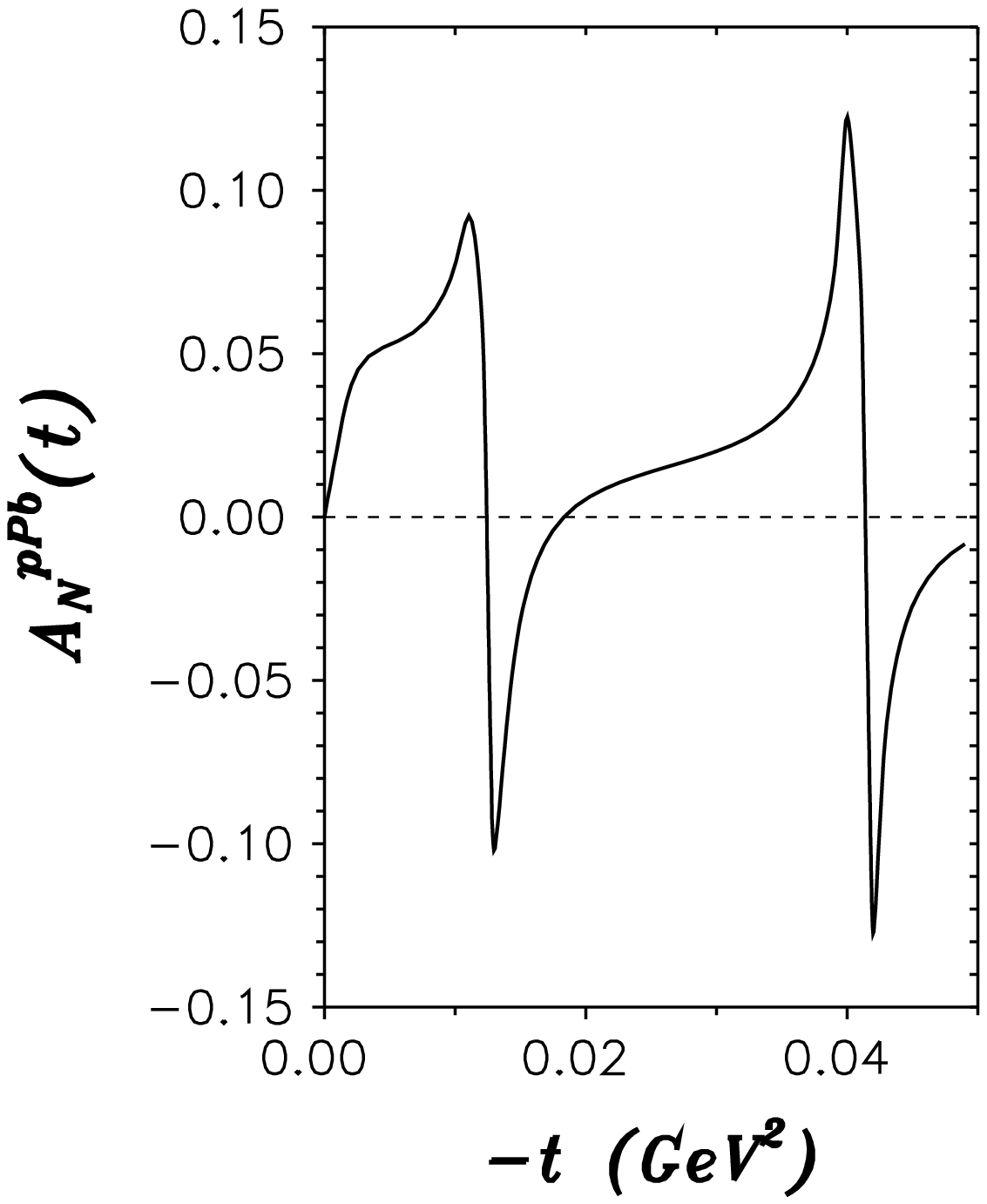}
\begin{center}
\vspace{8cm}
\parbox{13cm}
{\caption[Delta]
{\it Asymmetry in polarised proton scattering on
carbon and lead, calculated using (\ref{10}).
The experimental point for $pC$ scattering is
from \cite{e704-c}.}
\label{fig1}}
\end{center}
\end{figure}

In the colliding mode heavy nuclei have $1/Z$ smaller scattering
angle than protons at the same momentum transfer. Therefore,
they cannot be detected in the traditional CNI region of $t \sim 10^{-3}\ 
GeV^2$. As soon as a much stronger CNI asymmetry 
is expected in the vicinity of the
first diffractive minimum in the differential cross section,
it might be feasible to do measurements 
in this range of momentum transfer 
with the $p-A$ collider at RHIC.
The maximum of the
asymmetry for carbon, $A_N^{pC}(t)\approx 0.25$ is expected at $|t|=0.078\
 GeV^2$, what corresponds to the same scattering angle as in
$pp$ elastic scattering at $|t|\approx 2\times 10^{-3}\ GeV^2$.
These measurements can be performed with a fixed carbon target
as well.

The differential cross section of 
$p-^4He$ elastic scattering
exhibits a well developed minimum at $|t| 
\approx 0.2\ GeV^2$ 
%(see Fig.~4).
%\begin{figure}[tbh]
%\special{psfile=he.ps angle=0. voffset=-280. hoffset=90.
%hscale=40. vscale=40.}
%\begin{center}
%\vspace{7.5cm}
%\parbox{13cm}
% {\caption[Delta]
% {\it Data \cite{he} for the elastic $p- ^{4}He$ differential cross section
%at $400\ GeV$.}
%\label{fig1}}
%\end{center}
%\end{figure}
%
We expect a large asymmetry in the vicinity of the dip \cite{k-bnl},
which can be reliably calculated. The scattering angle of the $^4He$ is
the same as in $pp$ scattering at $|t|\approx 0.05\ GeV^2$, which is 
easy to measure in the whole RHIC energy range. An additional
advantage of $^4He$ is a lack of excitations.

\vspace{1cm}

{\large\bf\boldmath 4. Polarimetry with
elastic
$pp$ scattering at large
$|t|$}\\

I addition to the CNI region of very small $t$
the single asymmetry in $pp$ elastic scattering 
is known to be quite large and
nearly energy independent at large 
$|t|
\sim 1 - 2\ GeV^2$ \cite{snyder,fidecaro1,fidecaro2}
(see Fig.~4). 
\begin{figure}[tbh]
\includegraphics{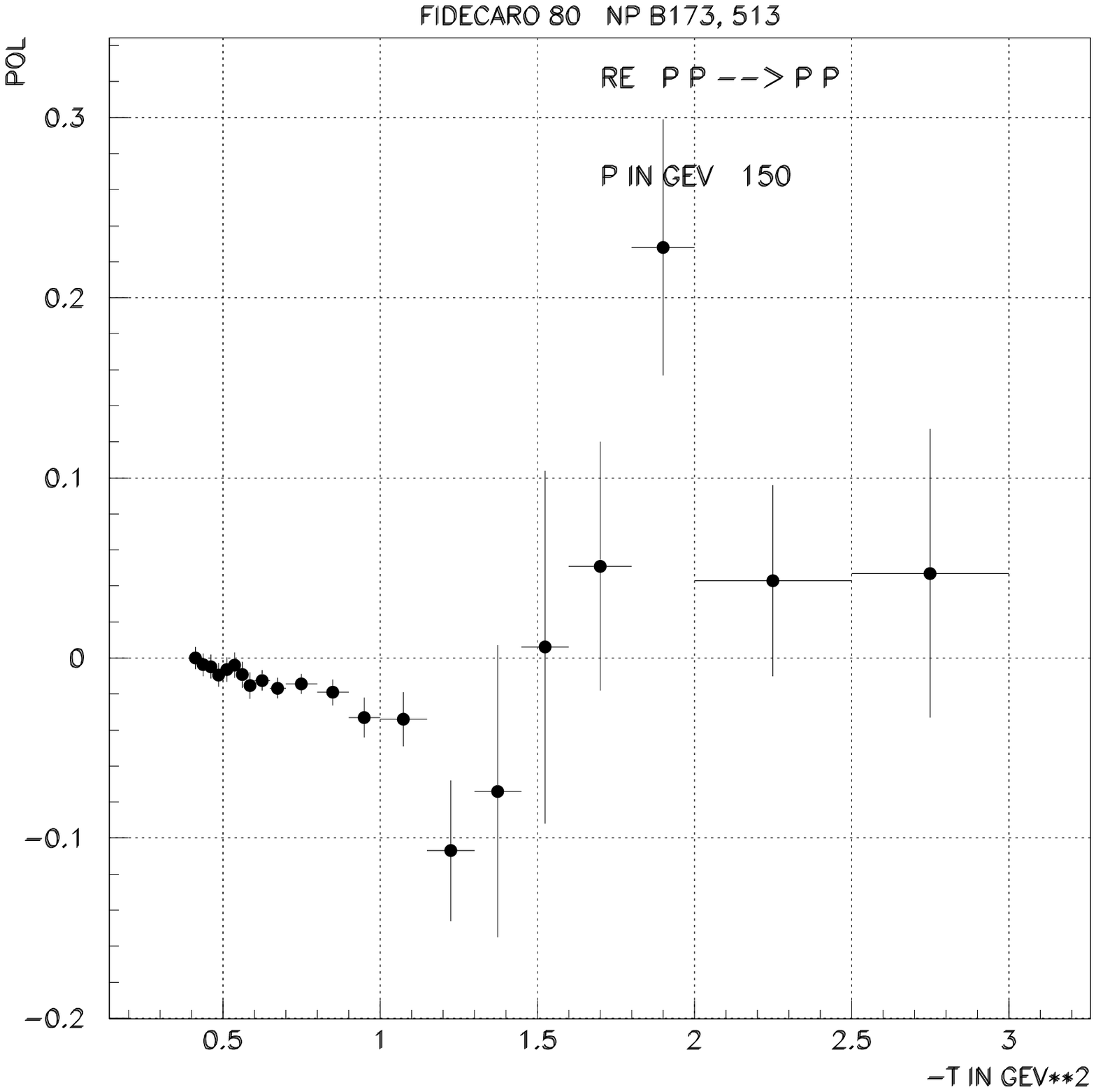}
\includegraphics{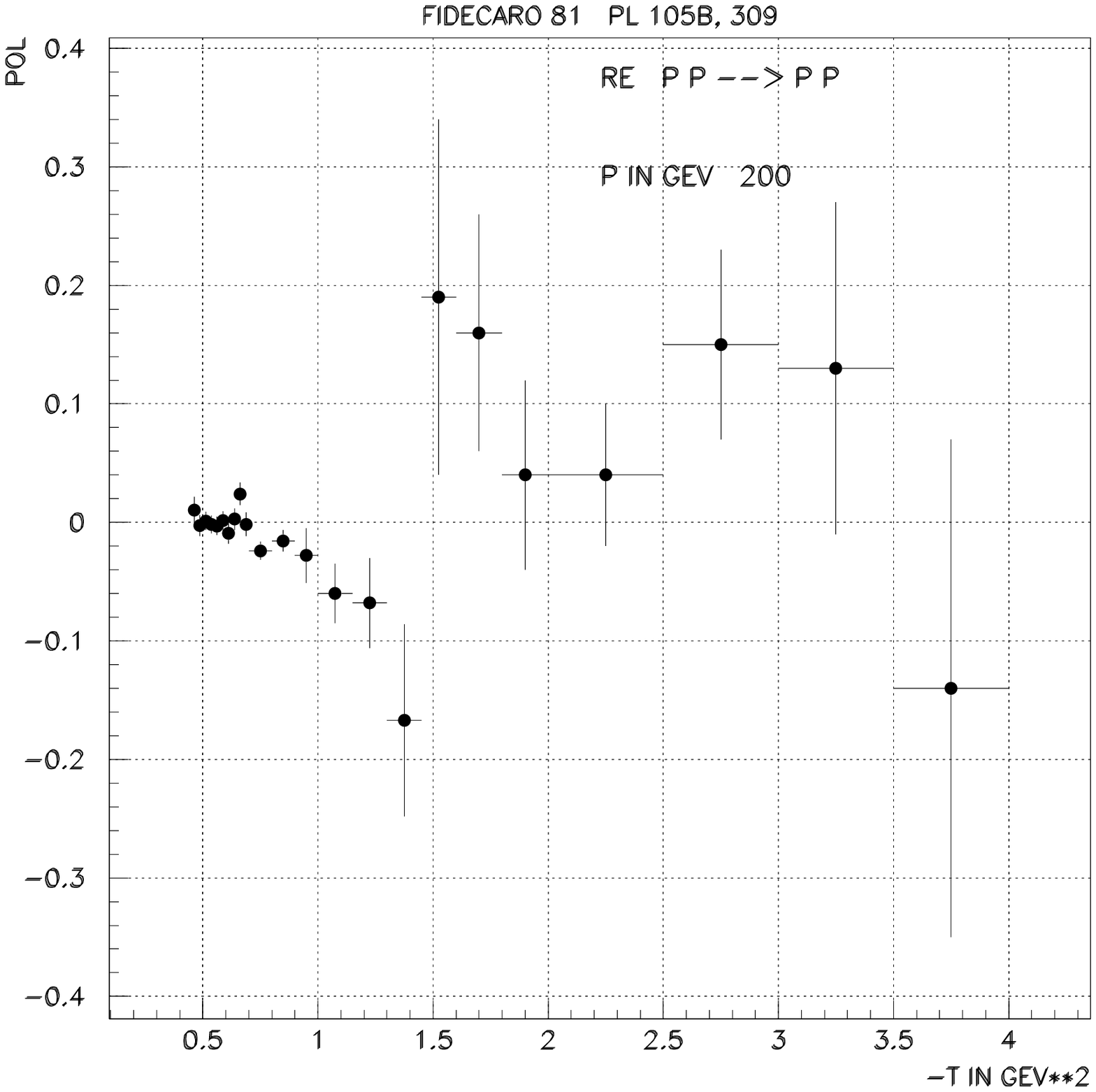}
\includegraphics{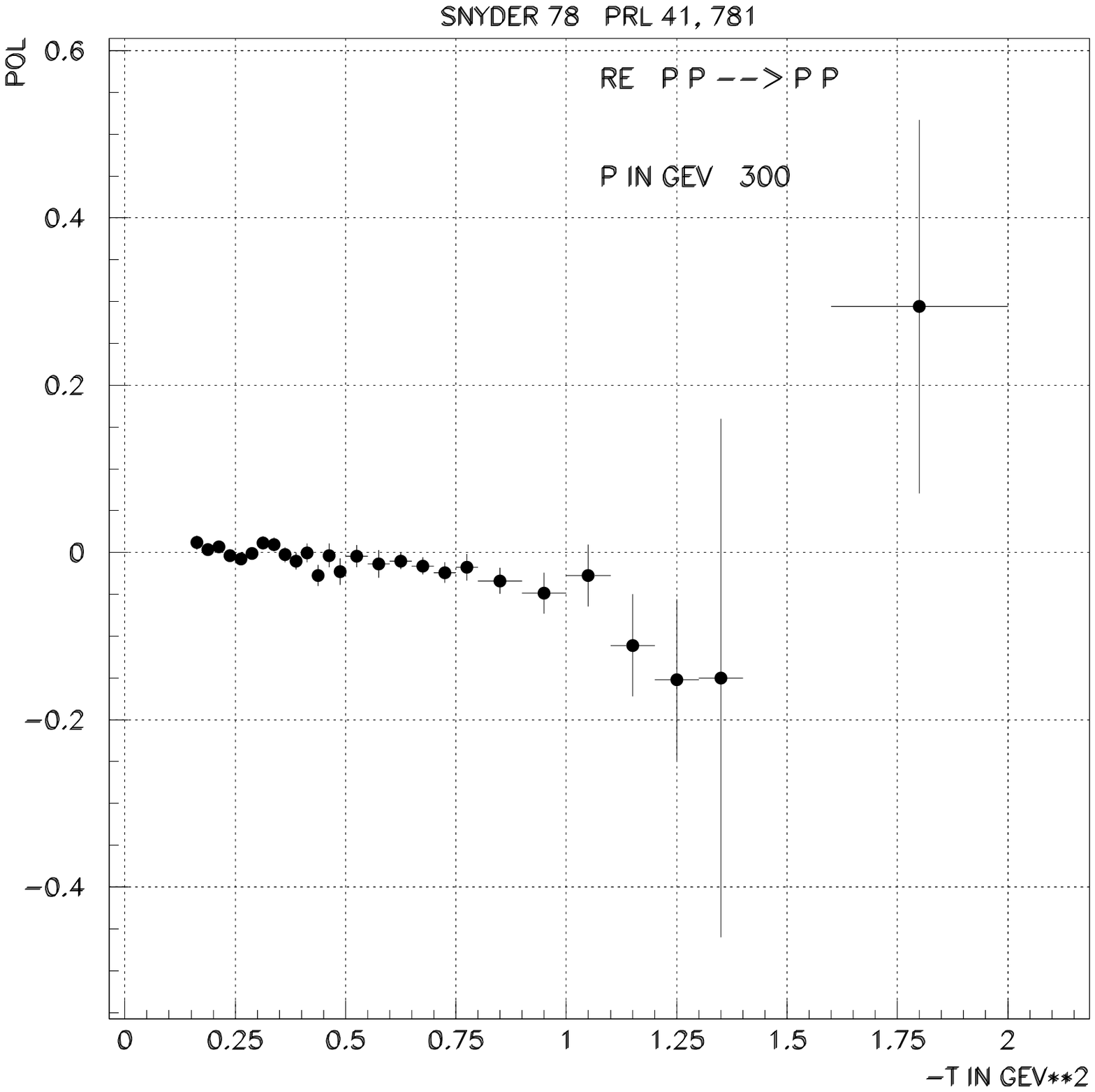}
\begin{center}
\vspace{5cm}
\parbox{13cm}
{\caption[Delta]
{\it Single asymmetry in polarised $pp$ elastic
scattering at $p_{lab} = 150,\ 200$ and $300\ GeV/c$ 
\cite{snyder,fidecaro1,fidecaro2}.}
\label{fig1}}
\end{center}
\end{figure}
This is related
to the dip structure of the differential
elastic $pp$
cross section in a similar way as we have seen above for
nuclear targets. 
The imaginary part of the non-flip part of the amplitude
is
unavoidably small near the point $t_0$ 
where it changes sign, particularly,
as small as the spin-flip part. Due to this fact
$A_N(t)$ is large and changes sign at
$t=t_0$. It is true at any energy, despite
the
decreasing energy dependence of the
spin-flip
amplitude.

To make use of this effect
for polarimetry one needs to know the
analysing
power $A_N(s,t)$. The available
data shown in Fig.~4 allow
only a crude evaluation of the
beam
polarisation. For a precise polarimetry
the
analysing power is to be measured with a
high
accuracy, what can be done in the
same
experiment. One should take the advantage
of
equality between the asymmetry of elastic
scattering
with a polarised beam and the polarisation of
final
protons with an unpolarised
beam. Then, one can measure
the
polarisation of recoil protons
instead of $A_N(s,t)$. The
kinetic
energy of the recoil proton
in the rest frame of the target is
quite
low,
\beq
E_{kin} = \frac{|t|}{2m_p}\
.
\label{15}
\eeq
In this energy range
$E_{kin}\approx
600\ MeV$ spin effects are known to be
quite
strong. One can use a standard carbon
polarimeter,
which can be calibrated to a high precision at
any
of available polarised beam facilities 
in this energy range
(e.g. at IUCF).

Thus, this method of polarimetry 
includes two
stages (provided that a calibrated
low-energy
carbon polarimeter is available). First, one
has
to calibrate the polarimeter with
an unpolarised
proton beam, {\it i.e.}
to
measure the recoil proton polarisation $P(s,t)
=
A_N(s,t)$ at $|t| \sim 1 - 1.5\ GeV^2$
at
different energies. Then one can remove
the recoil proton polarimeter and
measure the 
left-right asymmetry with a polarised
beam.
This asymmetry divided by $A_N$ gives the
beam
polarisation.

One can also use a fixed nuclear target, since
no depolarisation of the recoil proton in
nuclear matter is expected. However, the inelastic 
background is larger because of the broadening of
the recoil proton angle by Fermi motion.

Note that the results of such measurement would allow
to calibrate the CNI polarimeter, {\it i.e.}
to
eliminate the uncertainty of unknown
hadronic
spin-flip. One can use both small and large
$|t|$
polarimeters within the same experiment on
elastic
$pp$ scattering. This would provide a double
check
of the
results.

It worth also noting 
that with a polarised target one does
not
need the second scattering and the
low-energy
carbon polarimeter.

\vspace{1cm}

{\large\bf 5. Summary}\\

Comparison of three possible types of
polarimeters
for the RHIC polarised beams led us to
a
conclusion that probably
the
best is one which uses elastic
$pp$ and $pA$ 
scattering. Small $|t|$ scattering on
an
unpolarised target is based
on
theoretically predicted CNI analysing power,
which
has a relative uncertainty within $10\%$.
With nuclear targets one can perform the measurements at
larger $|t|$ where the differential 
cross section has a minimum. In this region the
CNI leads to dramatic polarisation effects.

Elastic $pp$ scattering at large $|t| \sim 1 -
1.5\
GeV^2$ on a fixed target can be calibrated using
a
second scattering of the recoil low-energy
proton.
This method, if
realistic\footnote{preliminary
evaluation shows \cite{wlodek} that such
a
polarimeter is feasible}, is
potentially
able to provide a most accurate, uncertainty
free
measurement of the beam polarisation. It
seems
that most effective is usage of
both
methods, which can be combined within the
same
experiment on elastic $pp$
scattering.

\vspace{0.5cm}

\noi
{\bf Acknowledgements} This work was initiated by Larry Trueman,
who invited me to the  Workshop on CNI polarimetry at RIKEN
Research Center, BNL. I appreciate very much the stimulating
and helpful discussions with Nigel Buttimore, Gerry Bunce, 
Wlodek Guryn, Elliot Leader, Yousef Makdisi, Jacques Soffer, and
Larry Trueman. The reading of the manuscript and many
improving corrections done by Nigel Buttimore are very much
appreciated.

This works was partially supported by the RIKEN
Research Center, BNL and INTAS grant 93-0239ext.


\begin{thebibliography}{10}

\bibitem{yousef} Y.~Makdisi, Talk given at the workshop on
CNI polarimetry, RIKEN Research Center, Brookhaven National
Laboratory, July 20 - August 23, 1997

\bibitem{e704-pi} The E704 Coll., D.L.~Adams et al.,
Phys. Lett. {\bf 264B} (1991) 462

\bibitem{cronin} J.~Cronin et al., Phys. Rev. Lett.
{\bf 31} (1973) 426

\bibitem{dijet}
B.Z.~Kopeliovich,
Phys. Lett. {\bf
B343} (1995) 387

\bibitem{antreasyan} D. Antreasyan et al.,
Phys. Rev. {\bf D19} (1979) 764

\bibitem{ws} H.~De~Vries, C.W.~De Jager and C.~De~Vries, Atomic Data
and Nucl. Data Tables, {\bf 36} (1987) 469

\bibitem{povh} J.~H\"ufner and B.~Povh, Phys.Rev.Lett.{\bf 58} (1987) 1612;
Phys.Lett. {\bf B245} (1990) 653

\bibitem{kl} B.Z.~Kopeliovich and L.I.~Lapidus, Yad. Fiz. {\bf 19}
(1974) 218 [Sov. J. Nucl. Phys. {\bf 19} (1974) 114]

\bibitem{bgl} N.H.~Buttimore, E.~Gotsman and E.~Leader,
Phys. Rev. {\bf D 18} (1978) 694

\bibitem{kz} B.Z.~Kopeliovich and B.G.~Zakharov, Phys. Lett.
{\bf B226} (1989) 156

\bibitem{nigel} N.~Akchurin, N.H.~Buttimore and
A.~Penzo, Phys. Rev. {\bf D 51} (1995) 3944

\bibitem{larry} T.L.~Trueman,  hep-ph/9610316; hep-ph/9610429

\bibitem{e704} The E704 Coll., N.~Akchurin et al.,
Phys. Rev. {\bf D 48} (1993) 3026

\bibitem{workshop} N.H.~Buttimore, B.Z.~Kopeliovich, E.~Leader,
J.~Soffer and T.L.~Trueman, paper in preparation

\bibitem{z} B.G.~Zakharov, Yar. Fiz. {\bf 49} (1989) 1386
[Sov. J. Nucl. Phys. {\bf 49} (1989) 860]

\bibitem{k80} B.Z.~Kopeliovich, 'Polarisation Phenomena at High Energies 
and Low Momentum Transfer', in Proc. of the 
Symposium "Spin in high-energy physics", Dubna,1982, p.97

\bibitem{ira} Yu.M.~Kazarinov, I.K.~Potashnikova, B.A.~Khachaturov, 
A.A.~Derevshchikov, preprint 
JINR-P1-85-426, Jun 1985 (in Russian).

\bibitem{itep} K.G.~Boreskov, A.A.~Grigoryan, A.B.~Kaidalov
and I.I.~Levintov, Yad. Fiz. {\bf 27} (1977) 813
[Sov. J. Nucl. Phys. {\bf 27} (1977) 432]

\bibitem{gol} S.V.~Goloskokov, Yad. Fiz. {\bf 39} (1984) 913

\bibitem{k-bnl} B.Z.~Kopeliovich, paper in preparation

\bibitem{schiz} A.M.~Schiz et al., Phys. Rev. {\bf D 21} (1980) 3010

\bibitem{nigel} N.H.~Buttimore, private communication

\bibitem{e704-c} The E704 Coll., N.~Akchurin et al.,
Phys. Lett. {\bf B229} (1989) 299

\bibitem{he} E. Jenkins et al., Phys.Rev.{\bf D 23} (1981) 1895 

\bibitem{snyder} J.~Snyder et al., Phys. Rev. Lett. {\bf 41}
 (1978) 781; {\it ibid} {\bf 41} (1978) 1256

\bibitem{fidecaro1} M. Fidecaro et al.,  Nucl. Phys.{\bf B173}
(1980) 513

\bibitem{fidecaro2}  M. Fidecaro et al., Phys. Lett. {\bf 229B}
(1989) 299

\bibitem{wlodek} W.~Guryn, private communication

\end{thebibliography}
\end{document}